\documentclass[aps,prl,reprint,superscriptaddress]{revtex4-2}

\usepackage[czech,english]{babel}

\usepackage{graphicx}

\usepackage{amsmath}
\usepackage{amsfonts}
\usepackage{amssymb}
\usepackage{amsthm}

\usepackage{xcolor}

\begin{document}

\title{Numerical study of vortices within a background vortex}

\author{L.~Xu}
\email{Contact author: lxu@ncat.edu}
\affiliation{Department of Mathematics and Statistics, North Carolina A\&T State University, Greensboro, 27411, NC, USA}

\date{\today}


\begin{abstract}
 This study aims to exploit the analogy of vortex dynamics in a 2D ideal fluid and 2D non-neutral plasma. Numerical simulations using contour dynamics with adaptive refinement are conducted to study the dynamics of one or more vortices within a background vortex, and results are compared with experiments using magnetized non-neutral electron plasmas confined in a Penning-Malmberg trap. This comparison is based on the isomorphism between the 2D Euler equations for ideal fluid and the 2D drift-Poisson equations for non-neutral plasma, where the the fluid vorticity and stream function correspond to the plasma charge density and electric potential. Agreement and discrepancy between simulation and experiment are observed and discussed. 
\end{abstract}

\maketitle

\textit{Introduction} --- The vortex dynamics of a  2D ideal fluid is ubiquitous in nature, for example, turbulence and tornado, where large-scale coherent structures emerge from smaller scale vortex structures. 
The motion of vortex in an ideal fluid is governed by the 2D Euler equations~\eqref{eqn:fluid}, 
\begin{equation}
\label{eqn:fluid}
\frac{\partial\omega}{\partial t} + 
{\bf u} \cdot \nabla\omega = 0,
\quad {\bf u} = \nabla^\perp\psi,
\quad -\nabla^2\psi = \omega,
\end{equation}
where $\omega(x,y,t)$ is the fluid vorticity,
${\bf u} = {\bf u}(x,y,t)$ is the velocity,
and
$\psi(x,y,t)$ is the stream function.
It is well-known that these equations are isomorphic to the 2D drift-Poisson equations \eqref{eqn:plasma} for non-neutral plasma,
\begin{equation}
\label{eqn:plasma}
\frac{\partial n}{\partial t} +
{\bf u} \cdot \nabla n = 0, \quad
{\bf u} = \nabla^\perp\phi, \quad
\nabla^2\phi = n,
\end{equation}
where
$n(x,y,t)$ is the plasma density,  
${\bf u}(x,y,t)$ is the drift velocity,
$\phi(x,y,t)$ is the electric potential.
For reference,
the electric field is ${\bf E} = -\nabla\phi$.
The equations are supplemented by an 
initial vorticity $\omega_0(x,y)$ or plasma density $n_0(x,y)$.

Non-neutral plasma has been a convenient tool to explore vortex structures in ideal fluids over past decades~\cite{levy1965diocotron,briggs1970role,driscoll1990experiments}.
Durkin and Fajans \cite{durkin2000experimental} placed a pointlike plasma clump/vortex  
within a diffuse circular plasma clump/background vortex, and they found that the pointlike vortex induces a wave on the disk perimeter, 
causing wave breaking and filamentation. The deformation later evolves into an empty hole in the disk which 
amounts to an antivortex. 
When there are multiple pointlike plasma clumps/vortices within a diffuse background vortex, they evolve into a symmetric state, which is
in accordance with Onsager's theory 
of equilibrium statistical mechanics~\cite{onsager1949statistical}. 
Jin and Dubin \cite{jin2001point} looked into the distorted perimeter of the background vortex and studied the wave breaking time analytically. In their work, the finite-size plasma vortex is treated as a point vortex
which interacts with the Kevin waves traveling on the boundary of the background vortex patch. 
Once the wave breaking occurs, the dynamics of point vortex are the same as the dynamics of point vortex within a fixed cylindrical  boundary.
%
When there are multiple plasma vortices inside a background plasma vortex, they will evolve into a regular symmetric shape.
Schecter \cite{schecter2002two} studied the effect of the background vortex using plasma experiments and found that it can 
cool down chaotic motions.  
They also investigated the effect of the background vortex numerically using positive and negative point vortices in axisymmetric background vorticity field. It is seen that perturbation on the background vorticity  causes the vortex inside to drift radially. 
%

In this work, we revisit the problem of one or more vortices within a background vortex \cite{durkin2000experimental} and examine the vortex dynamics by directly solving the Euler equations \eqref{eqn:fluid}. A primary goal is to compare results with those in experiments using non-neutral plasma and explore both agreement and discrepancy. 

\textit{Method and Problem} --- 
We assume a finite area vortex region with constant vorticity, and thus the method of contour dynamics \cite{zabusky1979contour, dritschel1989contour, schoemaker2003contour} is most suitable. 
The boundary of the vortex region is composed of a series of Lagrangian points. One evaluates the velocity \eqref{eqn:vel} on these Lagrangian points using Simpson's rule, and then moves them at the resulting velocity,
\begin{eqnarray}\label{eqn:vel}
\hskip -10pt
{\bf u}(\mathbf{z},t) = \frac{1}{4\pi}\sum_{j}[\omega]_{C_j}\oint_{C_j}\log \left( |\mathbf{z} - \mathbf{z}_j|^2 + \delta^2\right)~\frac{d\mathbf{z}_j}{d\alpha}~d\alpha. 
\end{eqnarray}
In the above formula, ${\bf u}(\mathbf{z},t)$ is the velocity at location $\mathbf{z}= (x,y)$ and time $t$, 
 $\delta$ is smoothing parameter to get rid of singularity, 
$C_j$ is the $j$th closed vortex contour, 
$ [\omega]_{C_j}$ is the constant vorticity at contour $j$, 
variable $\alpha\in[0,~1]$ is Lagrangian variable that is used to parameterize each contour. 
We using the panel-particle structure and adaptive point insertion based on techniques presented in~\cite{xu2023dynamics}.

\begin{figure}[htb]
\centering
\includegraphics[width=0.45\textwidth]{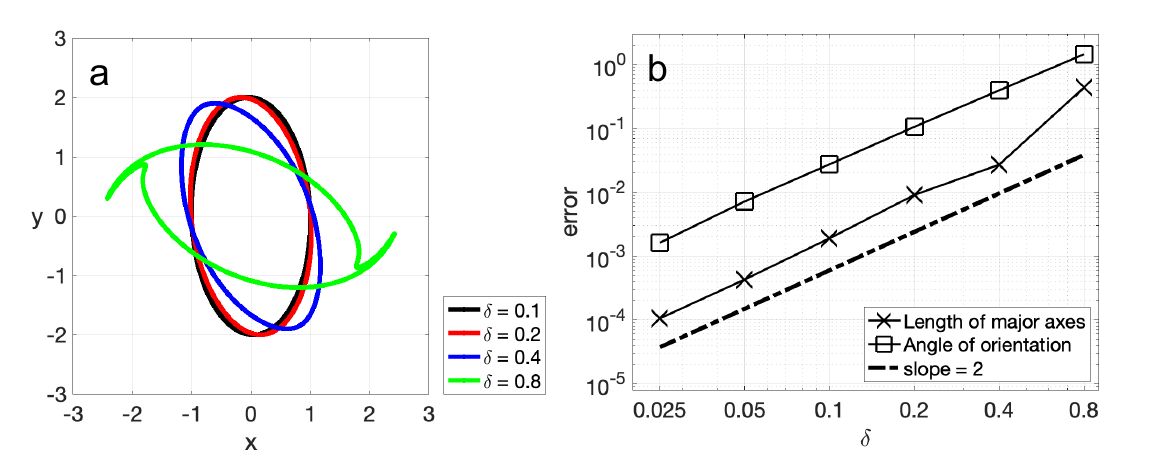}
\caption{Convergence in the smoothing parameter $\delta$ using elliptical vortex patch at the end of one whole rotation period. 
The patch has aspect ratio of 2:1 and a flat vorticity of 1.
(a) location of the vortex patch at $\delta=[0.1,~0.8]$ , 
(b) errors in the length of major axes and angle of orientation as a function $\delta \in [0.025, ~0.8]$.
}
\label{fig:figure-ellipse-free-conv}
\end{figure}

The convergence of the smoothing parameter $\delta$ is shown in Figure \ref{fig:figure-ellipse-free-conv} using an elliptic vortex patch~\cite{saffman1995vortex}. 
The elliptical vortex patch has aspect ratio of 2:1 and a flat vorticity of 1. 
Figure~\ref{fig:figure-ellipse-free-conv}a plots the vortex patch at four values of $\delta = 0.1, 0.2, 0.4, 0.8$ at the end of one whole period rotation. The patch remains elliptical except for the largest $\delta = 0.8$. The patch is closer to  its initial location at a smaller value of $\delta$. 
Figure~\ref{fig:figure-ellipse-free-conv}b shows associated errors in the length of major axes and angle of orientation as a function of $\delta$. It is seen that the errors in both measurements are decreasing as $\delta$ decreases, and a convergence order of $2$ is observed. We are using $\delta = 0.025$ in computation. 

\begin{figure}[htb]
\centering
\includegraphics[width=0.4\textwidth]{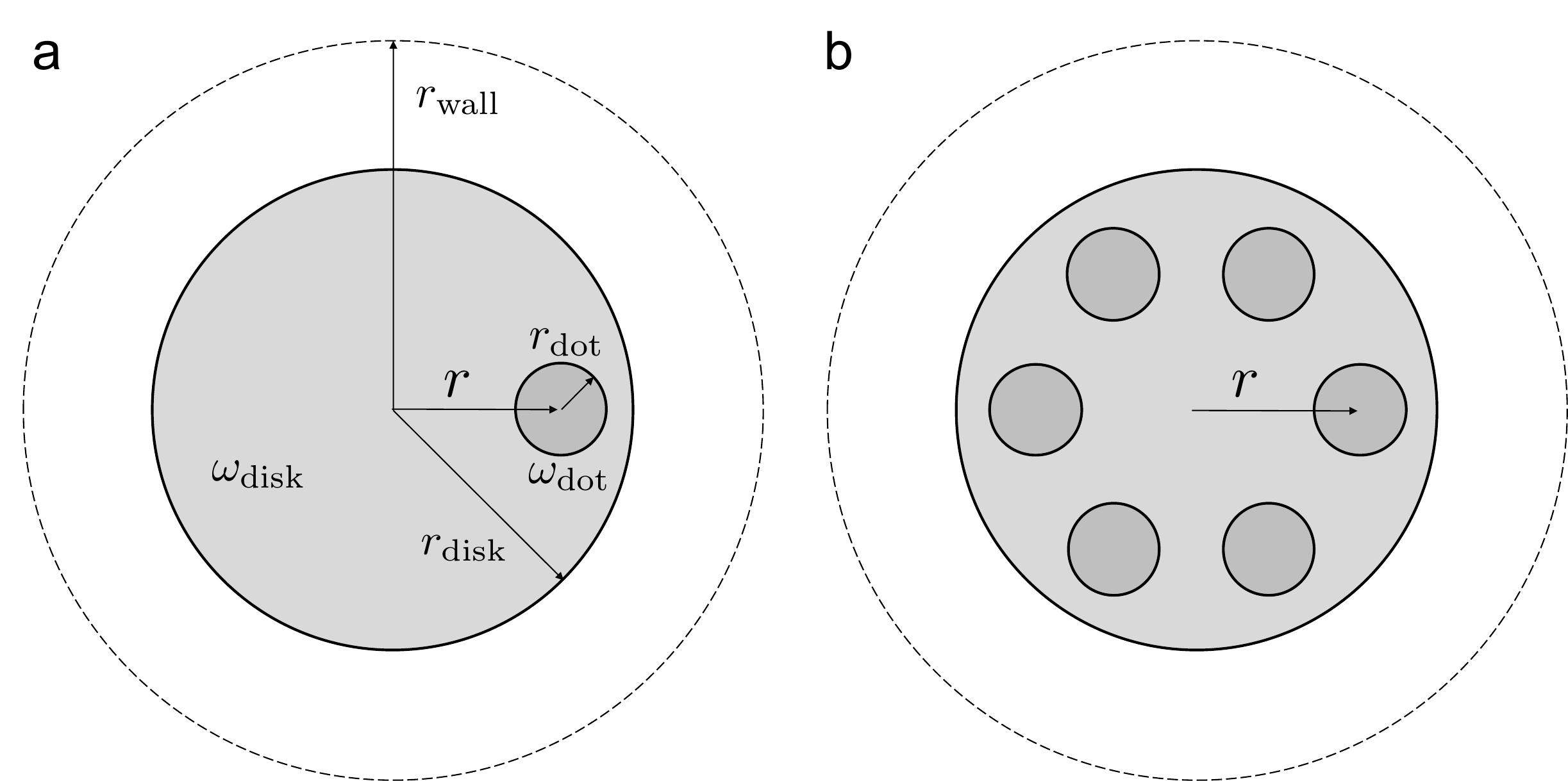}
\caption{
Problem setup. 
(a) a single vortex in a background vortex, 
(b) six symmetric vortices in a background vortex. 
}
\label{fig:figure-setup}
\end{figure}

Figure~\ref{fig:figure-setup} displays the problem setup following experiments of Durkin and Fajans~\cite{durkin2000experimental} (figure 1). Fig.\ref{fig:figure-setup}a is one single vortex in a background vortex. The  background vortex is centered at origin with radius $r_\text{disk}=0.6~{\rm cm}$ and vorticity $\omega_{\text{disk}} =3.396\times10^4~{\rm s}^{-1}$, the strong vortex inside is at radius $r_\text{dot}=0.059~{\rm cm}$ on $x$-axis 
with vorticity $\omega_{\text{dot}}=18.871\times10^4~{\rm s}^{-1}$.
The conversion between vorticity level and plasma density is stated in Amoretti \textit{et al.}~\cite{amoretti2001asymmetric}.
A circular wall is present at $r_\text{wall}=2~{\rm cm}$ and image points are used to ensure velocity normal to the wall vanishes. 
Fig.\ref{fig:figure-setup}b shows six symmetric vortices in a background vortex. 

\begin{figure}[ht!]
\centering
\includegraphics[width=0.5\textwidth]{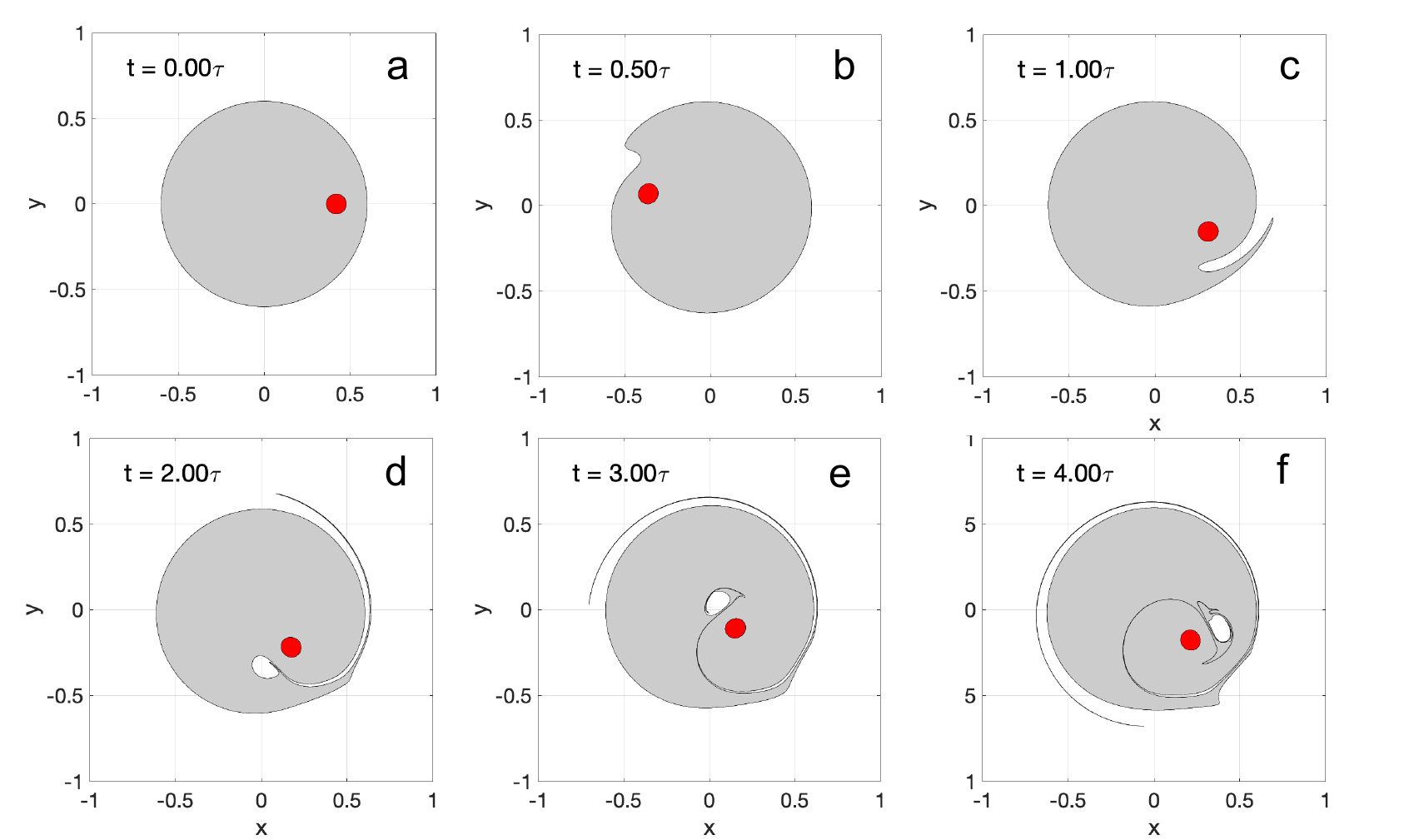}
\caption{
A strong vortex (red) within a weak background vortex (gray) at time $t = [0,~4\tau]$.
}
\label{fig:figure-fig1-color}
\end{figure}

\textit{Dynamics of a single vortex in a background vortex} --- Figure \ref{fig:figure-fig1-color} shows the dynamics of a strong vortex (red) in a weak background vortex (gray). 
Fig.\ref{fig:figure-fig1-color}a displays initial locations of contours at $t=0\tau$. The variable  $\tau  = 4\pi/\omega_\text{disk}$, which is analytic time when the background vortex rotates an angle of $\pi$.

At $t=0.5\tau$, the strong vortex travels clockwise. 
The perimeter of the background vortex deforms at places closest to the strong vortex. 
At $t=1.0\tau$, the strong vortex remains circular and keeps its clockwise excursion, however, gets closer to the origin.
The perimeter of the background vortex is further distorted and forms a narrow channel through which the non-rotational fluid outside is entrained. 
At $t=2.0\tau$, the entrained fluid forms a water bag or anti-vortex next to the strong vortex, and its size is comparable to the latter, therefore, an effective vortex dipole has formed. At this time, the fluid channel has become extremely narrow and the background vortex leaves a long thin filament behind. 
At $t=3.0\tau$, the dipole trajectory is curved owing to the effect of background vortex, and at the same time the strong vortex gets closer to the origin. The outside filament becomes longer and wraps around the background vortex.  
At $t=4.0\tau$, the dipole stays within the background vortex, the shape of the strong vortex 
remains almost circular. 
The excursion of the dipole leaves a complicated trace of filament, and this implies that vorticity of the background vortex patch is no longer flat and vorticity gradients are present.

\begin{figure}[ht!]
\centering
\includegraphics[width=0.5\textwidth]{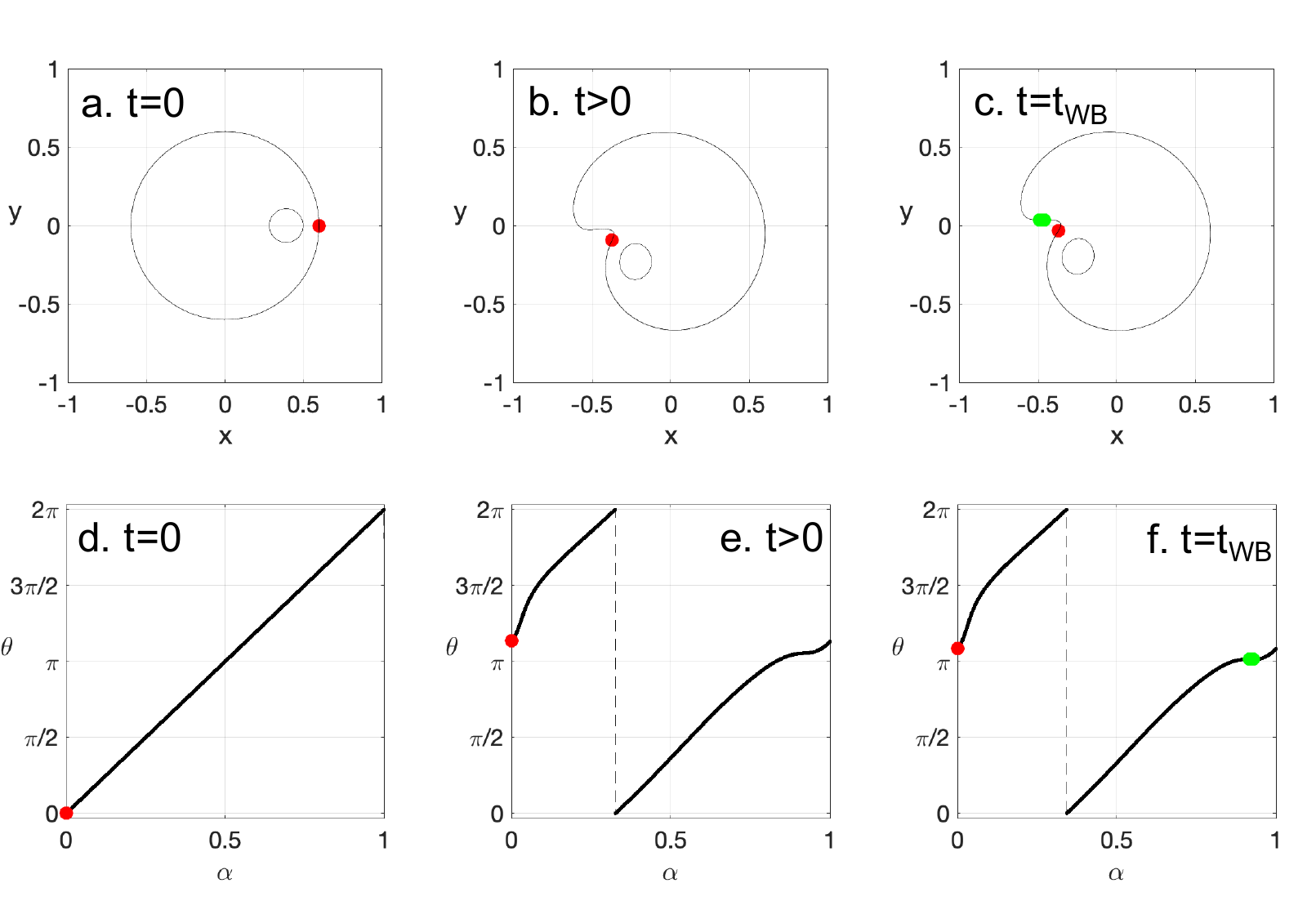}
\vskip -20pt
\caption{Illustration of wave breaking and definition of
the wave breaking time $t_{\rm wb}$. 
The top row (abc):  vortex contours at $t=0$, $t>0$, $t=t_{\text{wb}}$, 
the bottom row (def): the radial angle $\theta$ on the background contour is shown as a function of Lagrangian parameter $\alpha$ at same three times. 
Reference point $\alpha=0$ is marked in red, 
wave breaking location is marked in green. 
}
\label{fig:figure_twb_def}
\end{figure}

{\it Wave breaking for a single vortex in a
background vortex} --- The perimeter deformation of the background vortex is examined in terms of
wave breaking, the moment when the background contour forms a radial step. 
Figure~\ref{fig:figure_twb_def} illustrates the definition of wave breaking in the computation. 
Fig.\ref{fig:figure_twb_def}a shows two contours at $t=0$. 
Points on the background contour is ordered by a Lagrangian variable $\alpha\in[0,1)$ whose value increases counter-clockwise. 
The point at $(0.6,0)$ is marked in red and has $\alpha = 0$. 
We compute the radial angle $\theta$ on each point of the background contour, and  Fig.\ref{fig:figure_twb_def}d shows $\theta$ as a function of $\alpha$ in a linear plot. Point of $\alpha = 0$ is also labeled in red. It is seen that the plot is a straight line initially.

Fig.\ref{fig:figure_twb_def}(be) illustrate the contours at a positive time $t>0$. The red point moves together with the rotating background contour, and its radial angle  $\theta$ is increasing.
Fig.\ref{fig:figure_twb_def}(cf) show contours at the wave breaking time $t= t_{WB}$. It is seen that a radial step forms on the 
background contour, marked in green. 
The wave breaking time is recorded when the tangent line on the plot of 
$\theta$ vs.\ $\alpha$ attains a zero slope. 

\begin{figure}[ht!]
\centering
\includegraphics[width=0.4\textwidth]{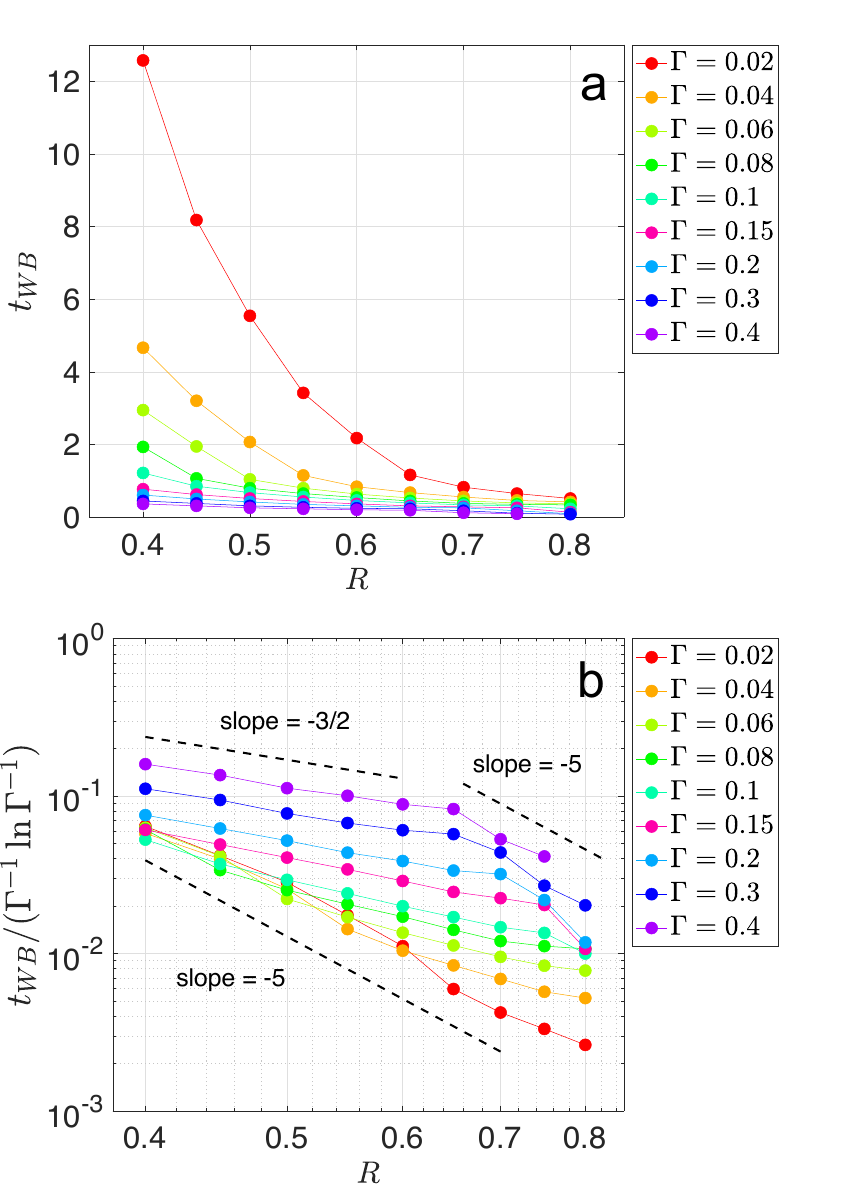}
\caption{
(a) Wave breaking time $t_{\rm WB}$ is plotted as a function of radial distance $R = r/r_{\text{disk}}$ in a linear plot, 
(b) scaled wave breaking time $t_{\rm WB}/\left(\Gamma^{-1}\ln \Gamma^{-1} \right)$ 
is plotted as a function of $R$ in the log-log plot,
circulation $\Gamma$ varies from $0.02$ to $0.4$, and $R$ ranges from $0.4$ to $0.8$.
}
\label{fig:figure-twb}
\end{figure}

Figure~\ref{fig:figure-twb} 
shows the dependence of the 
wave breaking time at varying circulation and initial location for the strong vortex inside. 
Fig.\ref{fig:figure-twb}a plots the wave breaking time  $t_{WB}$ as a function of the radial distance $R=r/r_{disk}$ of strong vortex in a linear scale. 
The circulation, $\Gamma = \omega_{\text{dot}}r^2_{\text{dot}}/ \omega_{\text{disk}}r^2_{\text{disk}}$, 
varies in $[0.02,~0.4]$ and $R$ ranges in $[0.4, ~0.8]$. 
The value of $\Gamma$ is adjusted by changing $r_{\text{dot}}$ and fixing $\omega_{\text{dot}}$, which is controlled the same way in experiments~\cite{durkin2000experimental}. 
%

In Fig.\ref{fig:figure-twb}a, 
it can be seen that the wave breaking occurs earlier at a 
larger radius $R$  at a given circulation $\Gamma$. 
This is because when these two contours are closer, it quickly induces distortion of the background perimeter. 
This kind of interaction becomes weaker when the vortex inside is away from the perimeter or when $R$ decreases. 
At a given distance $R$, 
the wave breaking time is delayed for a smaller circulation $\Gamma$, this is also due to the weaker interaction between these two contours. 
When $R \to 0.8$, all curves collapse together which agrees with experimental observation~\cite{durkin2000experimental} (figure 5a).

Figure~\ref{fig:figure-twb}b plots the scaled wave breaking time $t_{WB}/\left(\Gamma^{-1}\ln \Gamma^{-1} \right)$ as a function of radial distance $R$ in loglog scale. 
Two scenarios are observed depending on the value of $\Gamma$. 
When $\Gamma \in [0.02, 0.1]$, the curves collapse together at small values of $R$, and a reference line of slope $-5$ is plotted. However, these curves tend to level up and start to distinguish from each other as $R$ increases. These observations are also in great agreement with experiments~\cite{durkin2000experimental} (figure 5b).  
However, when $\Gamma \in [0.1, 0.4]$,  curves are parallel to each other, and 
experience two slopes. Taking $\Gamma = 0.4$, for example, the wave breaking time behaves as $t_{WB} \sim R^{-3/2}$  at $R<0.65$ and  
$t_{WB} \sim R^{-5}$ at $R>0.65$.
The transition of these two stages is marked by a sudden and faster decay in slopes, and such a decay occurs later at a smaller value of $\Gamma$. 
%

\begin{figure}[ht!]
\centering
\includegraphics[width=0.5\textwidth]{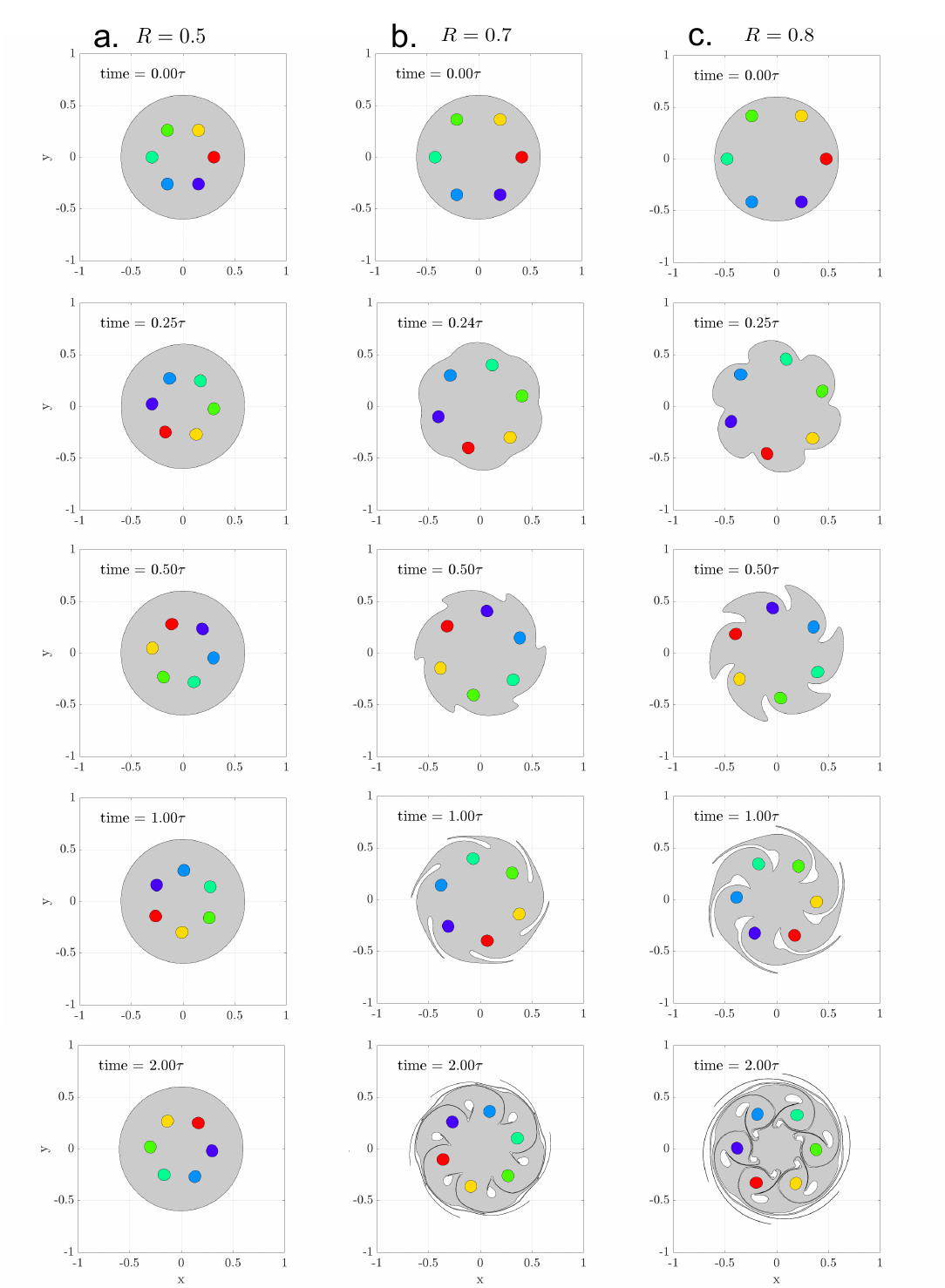}
\caption{
Dynamics  of six symmetric vortices within a background vortex at time $t=[0,~2\tau]$.
The radial distance of the vortices inside
is 
(a) $R = 0.5$,
(b) $R = 0.7$,
(c)  $R = 0.8$.
}
\label{fig:figure-multiple}
\end{figure}

\textit{Dynamics of six vortices within a background vortex}. 
Figure \ref{fig:figure-multiple} displays the time evolution of six symmetric vortices in a background vortex at $t=[0,~2\tau]$. 
The radial distance for the vortex inside is (a) $R=0.5$, (b) $R=0.7$, (c) $R=0.8$.
At $R = 0.5$, these six vortices rotate clockwise when time starts, and remain symmetric and distinct from each other at all times. There is no merger nor obvious interaction with the perimeter of the background vortex. 
%
At $R=0.7$, these six vortices are initially closer to the edge of the background vortex, which causes the deformation of the background perimeter. Wave breaking occurs during $t=0.5\tau$, and then fluid entrainment is seen around $t=1\tau$. 
At the end, each vortex inside is accompanied by a water bag or an anti-vortex,  which is composed of the entrained fluid. The antivortex is smaller compared to the neighboring vortex, and they together form an asymmetric dipole, see at $t=2\tau$. 
At $R=0.8$, these six vortices are initially further closer to the edge of the background vortex and the edge deforms immediately as time starts. 
Similarly, wave breaking occurs, fluid is entrained, and anti-vortices are formed inside.  
It is interesting to see that each vortex inside is now accompanied by two anti-vortices instead of one. There is a ring of six small anti-vortices closer to the origin and a ring of large anti-vortices close to the edge of the background vortex. The vortex inside and its two neighboring anti-vortices therefor form a tripole.
%

\begin{figure}[ht!]
\centering
\includegraphics[width=0.4\textwidth]{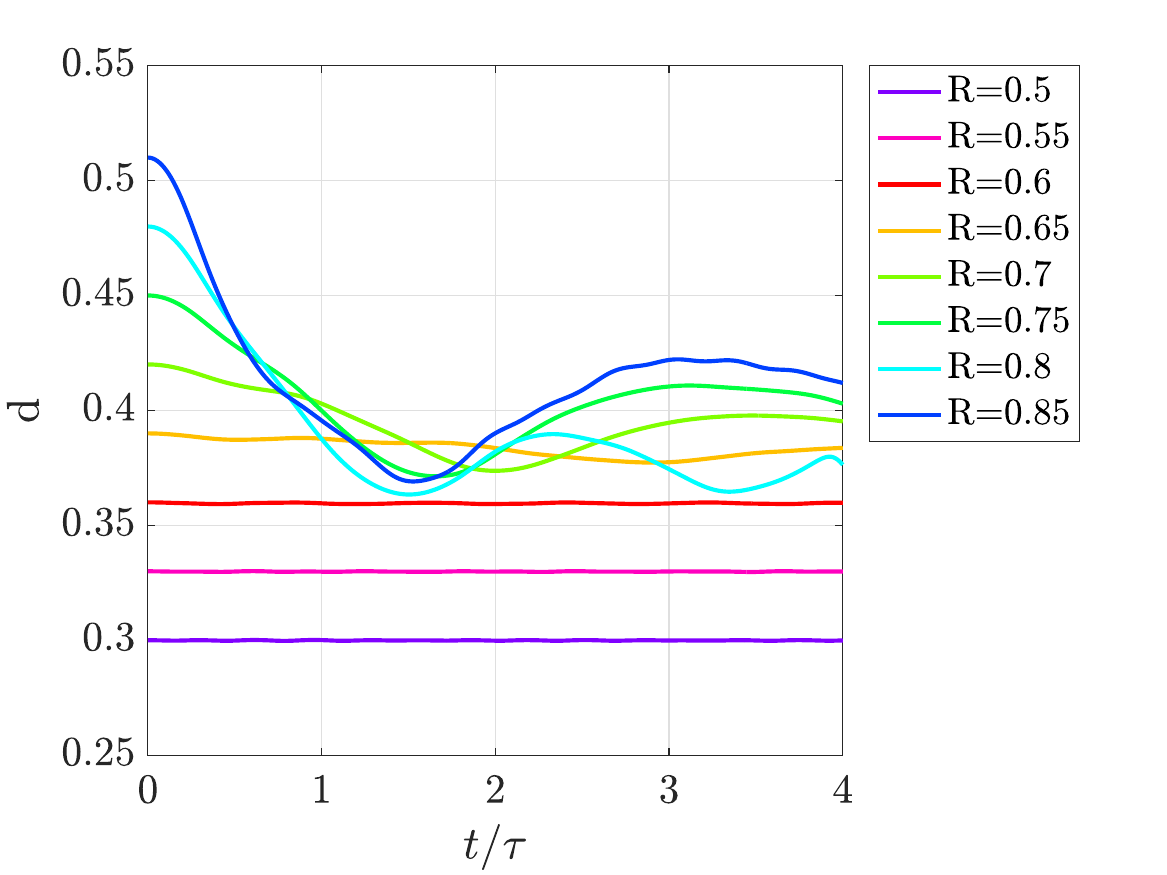}
\caption{
The time-dependent distance $d$ from the center of the strong vortex to the origin is plotted as a function of time $t/\tau$ for $R=[0.5,~0.85]$. 
}
\label{fig:figure-multiple-dist}
\end{figure}

The presence of a newly formed dipole and tripole will affect the motion of the inside vortices, thus their distance from the origin is time-dependent. 
Figure \ref{fig:figure-multiple-dist} shows the time-dependent distance from the center of the strong vortex $d$ as a function of time for $R$ ranging in $[0.5, ~0.85]$.  
The distance  $\displaystyle  d = |\mathbf{z}_c(t)|$, where $\displaystyle  \mathbf{z}_c(t) = \frac{\oint \mathbf{z}(s) ds}{\oint ds}$, $\mathbf{z}$ is the coordinate of points on a vortex contour inside, and $s$ is the arclength.
 
In Fig.\ref{fig:figure-multiple-dist}, two behaviors of $d$ are seen.  
At small initial value of $R\in[0.5,0.6]$, the distance $d$ remains approximately constant in time. 
At a larger value of $R\in[0.65, 0.85]$, the distance $d$ oscillates, first decreases and then goes up. The magnitude of the oscillation gets larger at a larger $R$.  The oscillating behavior is caused by the formation of dipole and triple inside.

\textit{Summary and Discussion}. 
Non-neutral electron plasmas have been used to study vortex dynamics for a long time based on the isomorphism between 2D drift-Poisson equations for plasma and 2D Euler equations for an ideal fluid. In this work, we have performed highly resolved numerical simulations on the vortex dynamics and aim to explore the analogy between these two systems. We revisit the popular experiments of Durkin and Fajans~\cite{durkin2000experimental} and compare results between the experiments and numerical simulations.
%
%
Our simulations display agreements with that of experiments, and at the same time, we are able to reveal more interesting details of vortex motions, such as wave break time scaling, formation of dipole and tripole,  filamentation, and fluid emanations, all of which in return yield new insights into non-neutral plasma dynamics. 

However, discrepancies between numerical simulations and experiments are also noted. 
%
We think that these discrepancies are attributed to three reasons.
The first one is in the problem setup. 
The simulation enables precise control of initial vorticity distribution and compact support of the vortex patches,  while electron density generated in experiments shows a bit of fluctuation in the patch area (color gray is not absolutely uniform) and a large gradient may be present at the edge.
The second one is in the limitation to the analogy between pure electron plasmas and two-dimensional inviscid fluids.
Peurrung and Fajans \cite{peurrung1993limitation} mentioned that
finite-length plasma experiences density smearing at short-length scales, while such smearing is absent in ideal fluid. 
The amplitude smaller than 0.1cm is entirely suppressed in experiments~\cite{durkin2000experimental}.
The third one lies in a substantial difference in the mechanism of these two isomorphic systems. 
The non-neutral plasmas will relax to thermodynamic equilibrium with symmetric patterns, according to Onsager's theory~\cite{onsager1949statistical, dubin1999trapped, krommes2002fundamental}. 
However, the vortex dynamics in an ideal fluid may relax into a complicated non-symmetric structure based on conservation of momentum~\cite{saffman1995vortex}. 

%

%

\begin{acknowledgments}
Fruitful discussions with Professor Robert Krasny is greatly appreciated. 
\end{acknowledgments}


\bibliography{refs}

\end{document}